\documentclass[preprint2,times,tighten]{aastex61}

\usepackage{epsfig}			
\usepackage{graphicx,color}		
\usepackage{amssymb}			
\usepackage{color}			
\usepackage{url}			
\usepackage{amsmath}			
\usepackage{rotating}			
\usepackage{float}			
\usepackage{textcomp}			
\usepackage{epstopdf}
\usepackage{dcolumn}
\usepackage{times}
\usepackage{tabularx}
\usepackage[english]{babel}

\shorttitle{Frequency-dependent damping in plumes and interplumes}
\shortauthors{Sudip Mandal et al.}

\begin{document}


\title{A Statistical Study On The Frequency-Dependent Damping of Slow-Mode Waves in Polar Plumes and Interplumes\\
}

\correspondingauthor{Sudip Mandal, email: sudip@iiap.res.in\\
$^{*}$now at: Astrophysics Research Centre, School of Mathematics and Physics, Queen's University Belfast, Belfast BT7 1NN, UK}

\author[0000-0002-7762-5629]{Sudip Mandal}
\affil{Indian Institute of Astrophysics, Koramangala, Bangalore 560034, India}

\author{S. Krishna Prasad$^{*}$}
\affil{Indian Institute of Astrophysics, Koramangala, Bangalore 560034, India}

\author{Dipankar Banerjee}
\affil{Indian Institute of Astrophysics, Koramangala, Bangalore 560034, India}
\affil{Center of Excellence in Space Sciences India, IISER Kolkata, Mohanpur 741246, West Bengal, India}

\begin{abstract}

We perform a statistical study on the frequency-dependent damping of slow waves propagating along polar plumes and interplumes in the solar corona. Analysis of a large sample of extreme ultraviolet (EUV) imaging data with high spatial and temporal resolutions obtained from AIA/SDO suggests an inverse power-law dependence of the damping length on the periodicity of slow waves (i.e., the shorter period oscillations exhibit longer damping lengths), in agreement with the previous case studies. Similar behavior is observed in both plume and interplume regions studied in AIA 171~\AA\ and AIA 193~\AA\ passbands. It is found that the short-period (2--6 min) waves are relatively more abundant than their long period (7--30 min) counterparts in contrast to the general belief that the polar regions are dominated by the longer-period slow waves. We also derived the slope of the power spectra ($\mathrm{\alpha}$, the power-law index) statistically to better understand the characteristics of turbulence present in the region. It is found that the $\mathrm{\alpha}$ values and their distributions are similar in both plume and interplume structures across the two AIA passbands. At the same time, the spread of these distributions also indicate the complexity of the underlying turbulence mechanism.

\end{abstract}

\section{Introduction}
 Slow magnetoacoustic waves (or simply, slow waves) are compressible magnetohydrodynamic (MHD) waves which typically propagate along the magnetic field lines in solar corona \citep{2000SoPh..193..139R}. These waves are found in a variety of structures across different layers of the solar atmosphere \citep{2007SoPh..246....3B}. However, in this paper, we are only interested in the coronal counterpart of these waves as seen in large-scale coronal structures such as polar plumes \citep{2016GMS...216..419B}. Polar plumes are the long and thin `ray' like structures which trace open magnetic field lines emerging from the unipolar coronal hole regions \citep{2015LRSP...12....7P}. One of the first detections of slow waves was reported by \citet{1998ApJ...501L.217D} where the authors had used EUV  imaging data to show propagating intensity disturbances (later interpreted as slow waves) along polar plumes. These waves are ubiquitous in the solar corona \citep{2012A&A...546A..50K} with their apparent speeds ranging from 50 to 150 $\mathrm{km~s^{-1}}$ \citep{2012SoPh..279..427K}. Such omnipresent nature, along with easy detectability, often make these slow waves an important tool for coronal seismology \citep{2003A&A...404L...1K,2011ApJ...727L..32V}.

Interestingly, these waves are subject to rapid damping while propagating along the coronal structures \citep{2002ApJ...580L..85O}. This damping is primarily controlled by the thermal conduction whereas the other dissipative mechanisms such as the viscosity and the radiation loss have a little role to play \citep{2003A&A...408..755D}. Furthermore, the wave damping is not uniform over the entire range of periodicities. In other words, the damping of slow waves has a frequency dependence. \citet{2012A&A...546A..50K} showed that waves with shorter periods are restricted close to the base of the coronal features while those with longer periods propagate higher up. \citet{2014A&A...568A..96G} have found two distinct height ranges (above the solar limb) with two contrasting damping rates. The frequency-dependent damping aspects in different coronal structures (on-disk fan loops and polar plumes/interplumes) have been explicitly shown  using a single observation by \citet{2014ApJ...789..118K}. The authors have demonstrated that the frequency dependence of the damping observed in the on-disk loop like structures and the polar plume/interplume regions is different and neither of the observed dependences could be successfully explained by a linear MHD wave theory. However, using 3D MHD simulations, \citet{2016ApJ...820...13M} have shown that the observed characteristics for the on-disk loops are still consistent with the thermal-conduction-dominated damping of slow waves. Yet, the peculiar frequency dependence found in the polar region structures remains unexplained. The earlier conclusions of \citet{2014ApJ...789..118K,2014A&A...568A..96G} were derived from isolated case studies based on single dataset and thus, a statistical study is needed to establish the robustness of those results.\\

In this work, we revisit this anomalous damping behavior of the slow waves in polar regions by analyzing a large number of datasets. In the following sections we describe the data and the analysis methods adopted, followed by the results and the conclusions.

\section{Data} 
In this study, we used high-resolution extreme ultraviolet (EUV) imaging data from the Atmospheric Imaging Assembly (AIA) \citep{2012SoPh..275...17L}, onboard Solar Dynamics Observatory (SDO) \citep{2012SoPh..275....3P} from two different passbands, namely, AIA 171~\AA\ and AIA 193~\AA. We selected a total of 62 datasets, each covering about 180 minutes duration, taken over a span of seven years from 2010--2017. All the datasets have a uniform cadence of 12~s with a spatial pixel scale $\approx$0.6$^{\prime\prime}$. Since our primary targets are polar plume/interplume structures, all of the selected data are associated with coronal holes. Consequently, the sample do not contain any datasets observed between the end of 2012 and the end of 2014, i.e., during the maximum of current solar cycle (cycle 24). This also keeps our data free from large-scale background structures, such as streamers. Besides, all the datasets were carefully checked with the $\mathrm{JHelioviewer}$ tool \citep{2013ascl.soft08016M} to exclude the possible presence of large and small-scale jets, missing frames, etc., in the selected data.\\

%

\begin{figure*}[!htb]
\centering
\includegraphics[width=0.98\textwidth]{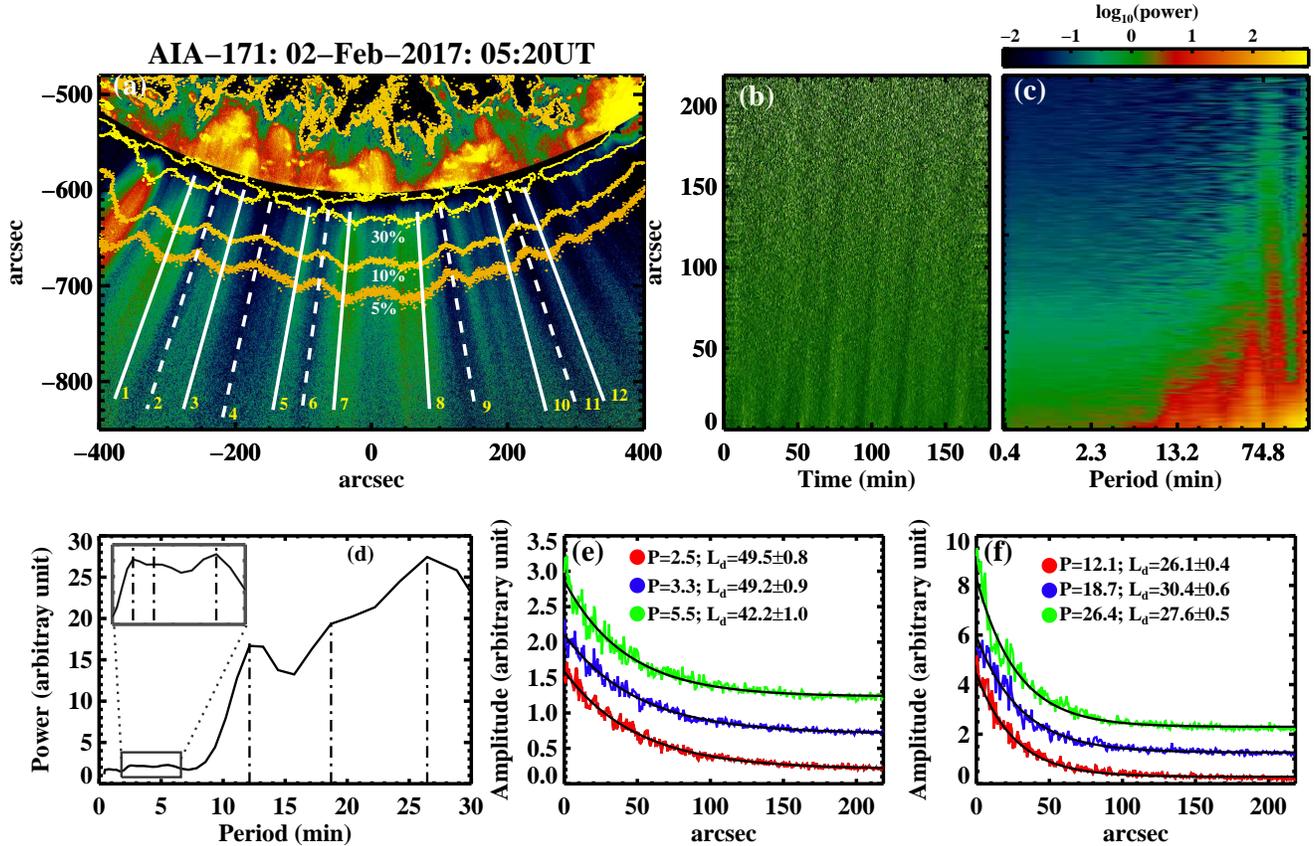}
\caption{Methodology adopted to analyze the frequency-dependent damping in slow waves: Panel (a) shows a representative image of our ROI which has been processed using a radial filter to enhance the polar features. Overplotted yellow contours denote different intensity levels (as \% of the on-disk intensity), obtained from the original image. Solid and dashed white lines represent the locations of the artificial slits used for generation of time-distance maps. An enhanced time-distance map, generated from slit 3, is shown in panel (b) whereas the corresponding period-distance map is shown in panel (c). Panel (d) displays a template power spectrum constructed from the period-distance map showing different periods present in the data. Panels (e-f) present the amplitude variation with distance for each of the detected periods. The black solid lines, in these plots, show the fitted exponential decay function to the data. The obtained damping lengths ($\mathrm{L_d}$) corresponding to individual periods ($\mathrm{P}$) are printed on the panels.} 
\label{context_image}
\end{figure*}

\section{Method}
\label{method}
The full-disk data are first reduced to a smaller regions of interest (ROIs) of size 800$^{\prime\prime}\times$400$^{\prime\prime}$ encompassing the polar region alone as shown in panel (a) of Figure~\ref{context_image}. The bright plume and the dark interplume structures are then detected by following different intensity contour peaks and valleys \citep{2011A&A...528L...4K,2014ApJ...793..117S}, respectively, as shown in the Figure. The methodology adopted here to track the slow waves of different periods and study their damping, is similar to that described in \citet{2014ApJ...789..118K} and \citet{2016ApJ...820...13M}. The overall procedure can be summarized as follows. First, we generate time-distance maps by extracting data from artificial slits placed on top of the detected plume and interplume structures. The white solid and dashed lines marked on Figure~\ref{context_image}a represent these slits for plume and interplume structures, respectively. The width of each of the slits is kept uniform and wide ($\approx$20$^{\prime\prime}$) to improve the signal to noise ratio in the generated time-distance maps. Figure~\ref{context_image}b displays one such map obtained from slit 3, which has been further enhanced by subtracting (and then normalizing with) a background constructed from a running average of 300 frames (60 minutes) of the original data. Note that the bottom of the map corresponds to the limb. In this enhanced map, we readily identify the upwardly propagating slow waves as alternative slanted ridges. Such calssification of these intensity disturbances is primarily based upon the measurements of their (projected) phase speeds from the time-distance maps \citep{2011A&A...528L...4K}. A closer look at this map also reveals the simultaneous existence of multiple periods. This is particularly evident near the limb where we see many closely spaced ridges (shorter periods) as compared to a far away region where only sparsely spaced ridges (longer periods) are visible. 

In order to isolate the power associated with the individual periods, we convert the original time-distance maps into period-distance maps (Figure~\ref{context_image}c) by using a wavelet transformation \citep{1998BAMS...79...61T} of the time series at each spatial position. We note that the power at the longer periods is higher compared to that at the shorter periods and the power, at a given period, decreases as one moves away from the solar limb. A template power spectrum (Figure~\ref{context_image}d) is constructed from the bottom 10 pixels of the period-distance map, which is used to identify the peak oscillation periods present in the data. For each detected period, the variation of the amplitude (defined as the square root of power) along the slit length is extracted as shown in Figures \ref{context_image}e and \ref{context_image}f. Considering the cadence (12~s) and the duration (180 minutes) of individual datasets, we restrict our analysis to periods between 2 and 30 minutes. As can be seen, the oscillation amplitudes decrease rapidly with the distance. To measure the amplitude decay quantitatively, we fit the data with an exponential function (shown as solid black lines in the figure) defined as $\mathrm{A(x)=A_0 e^{-x/L_d}+C}$, where $\mathrm{L_d}$ is the damping length. The $\mathrm{L_d}$ value obtained from the fit, for each detected period, is noted. In this way, we obtained damping lengths corresponding to multiple periods detected in several plume and interplume structures across all the 62 datasets in both 171~\r{A} and 193~\r{A} channels.

\section{Analysis}
\begin{figure}[!htb]
\centering
\includegraphics[width=0.5\textwidth]{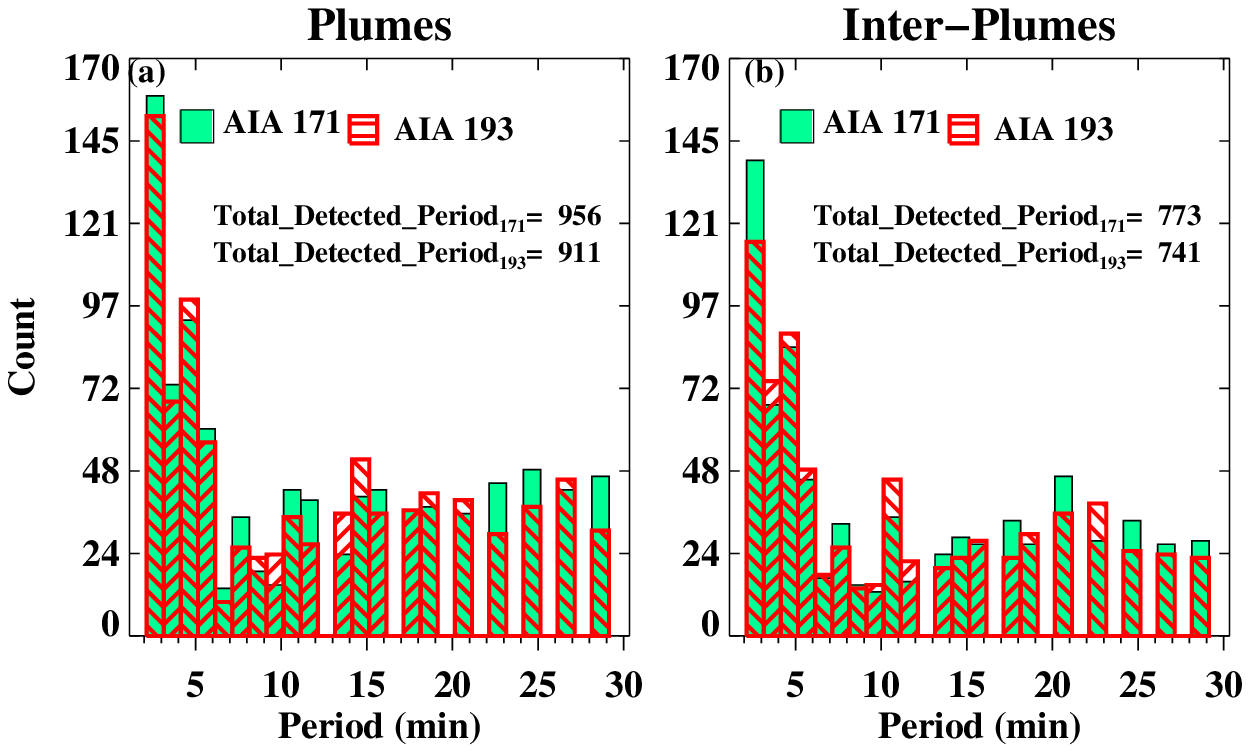}
\caption{Distribution of the detected periods in plume (a) and interplume (b) regions. The results from the 171~\r{A} and 193~\r{A} passbands are shown in color-filled and cross-hatched blocks, respectively.} 
\label{period_plot}
\end{figure}
\subsection{Period distribution}
In Figure~\ref{period_plot} we show a histogram of all the detected periods, in both plume and interplume structures observed in the two AIA passbands. The distribution is quite similar across different structures and passbands. The shorter periods (2--6 min) appear to be far more abundant than the longer periods ($>$7 min) which are more or less uniformly distributed. Although some previous studies \citet{2012A&A...546A..50K,2014A&A...568A..96G} reported the existence of both shorter and longer periods in the polar regions, it is generally believed that the long-period oscillations are more prevalent in these regions. So this behaviour is contradicting. Interestingly, upon carefully inspecting the detected periods from individual datasets, we found that the same set of shorter periods are observed in several plume/interplume structures whereas the longer periods observed in those structures are not necessarily the same. The latter are more or less evenly distributed across the period range between 7 and 30 min, producing such imbalance. Overall, about 41\% (43\%) of the detected periods in 171\r{A} (193\r{A}) passband are from shorter (2--6 min) periods when the data from both plume and interplume structures are combined. Since the observed oscillations are mainly due to propagating waves (the periodicity of which is determined by the source), it is possible that the shorter and the longer period oscillations are generated from a different source. For instance, the short-period oscillations could be a direct result of leakage of global $p$-modes \citep{2005ApJ...624L..61D, Calabro2013, 2015ApJ...812L..15K} or the chromospheric acoustic resonances \citep{2011ApJ...728...84B} or the acoustic cutoff effect involving impulvise disturbances \citep{2015ApJ...808..118C}., while the long-period oscillations are generated from transients (like spicules) in the lower atmosphere \citep{2015ApJ...809L..17J, 2041-8205-815-1-L16}. One must take these results into account while building a model (such as by \citet{2016ApJS..224...30Y}) to investigate the source of slow waves in polar regions.\\

\subsection{Frequency dependence}
\begin{figure*}[!htb]
\centering
\includegraphics[width=0.98\textwidth]{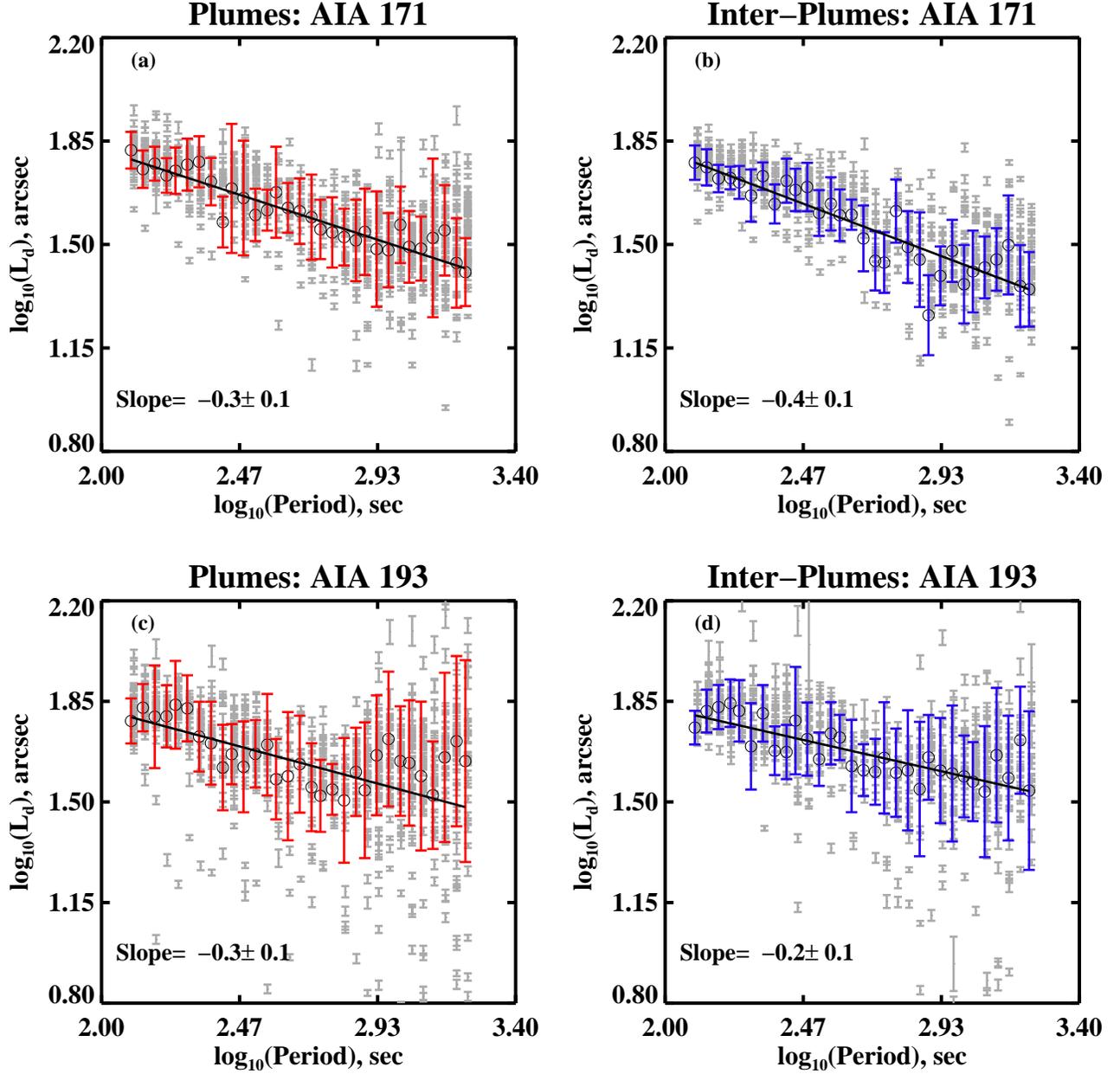}
\caption{Damping length as a function of oscillation period in log-log scale. The top (bottom) panels display the results for the AIA 171~\r{A} (193~\r{A}) passband, while the left (right) panels show the results for plumes (interplumes). In each plot, the grey symbols represent the individual measurements whereas the black open circles denote the most frequent value for a given period. See the text for more details.} 
\label{log_log}
\end{figure*}
The obtained damping lengths L$_d$, for all the detected periods $\mathrm{P}$ (see section~\ref{method}), are shown in a log-log plot in Figure~\ref{log_log} separately for plumes and interplumes and in the two AIA channels. The grey points in these plots represent the actual values with the corresponding uncertainties. As one may notice, the damping length values appear more scattered in 193~\r{A} passband as compared to that in 171~\r{A} passband. This is perhaps due to the lower signal-to-noise ratio in the 193~\r{A} passband. In fact, for a given passband, the scatter is relatively larger at longer periods. 
However, it is clear that the damping lengths become progressively shorter for longer periods. To quantitatively determine their dependence, we first construct a histogram of all the L$_d$ values at a period bin and consider the damping length corresponding to the peak of that histogram as the representative L$_d$ value at that period bin. Thus obtained L$_d$ values at each period bin are shown by black circles in Figure~\ref{log_log} (the errors are measured as the standard deviations of each distributions). These values are then fitted with a linear function (solid black line in the figure), the slope of which gives a measure of the dependence of damping length on the oscillation period. The obtained slope values are $-$0.3$\pm$0.1, and $-$0.4$\pm$0.1 for the plume and interplume structures, respectively, in the 171~\r{A} passband. The corresponding values in the 193~\r{A} passband are $-$0.3$\pm$0.1 and $-$0.2$\pm$0.1 for the plume and interplume structures. For comparison, the slopes obtained by \citet{2014ApJ...789..118K} are 
$-$0.3$\pm$0.1 and $-$0.4$\pm$0.1 in the 171~\r{A} and the 193~\r{A} passbands, respectively. Note that \citeauthor{2014ApJ...789..118K} did not obtain separate values for plume and interplume structures due to limited statistics. Nevertheless, the fact that the slopes obtained from a large statistics in the present study are not very different from that reported by \citeauthor{2014ApJ...789..118K} from a single case study, implies, the observed anomalous dependence of damping length on oscillation period is a general characteristic of slow waves in polar regions.
 
\begin{figure*}[!htb]
\centering
\includegraphics[width=0.98\textwidth]{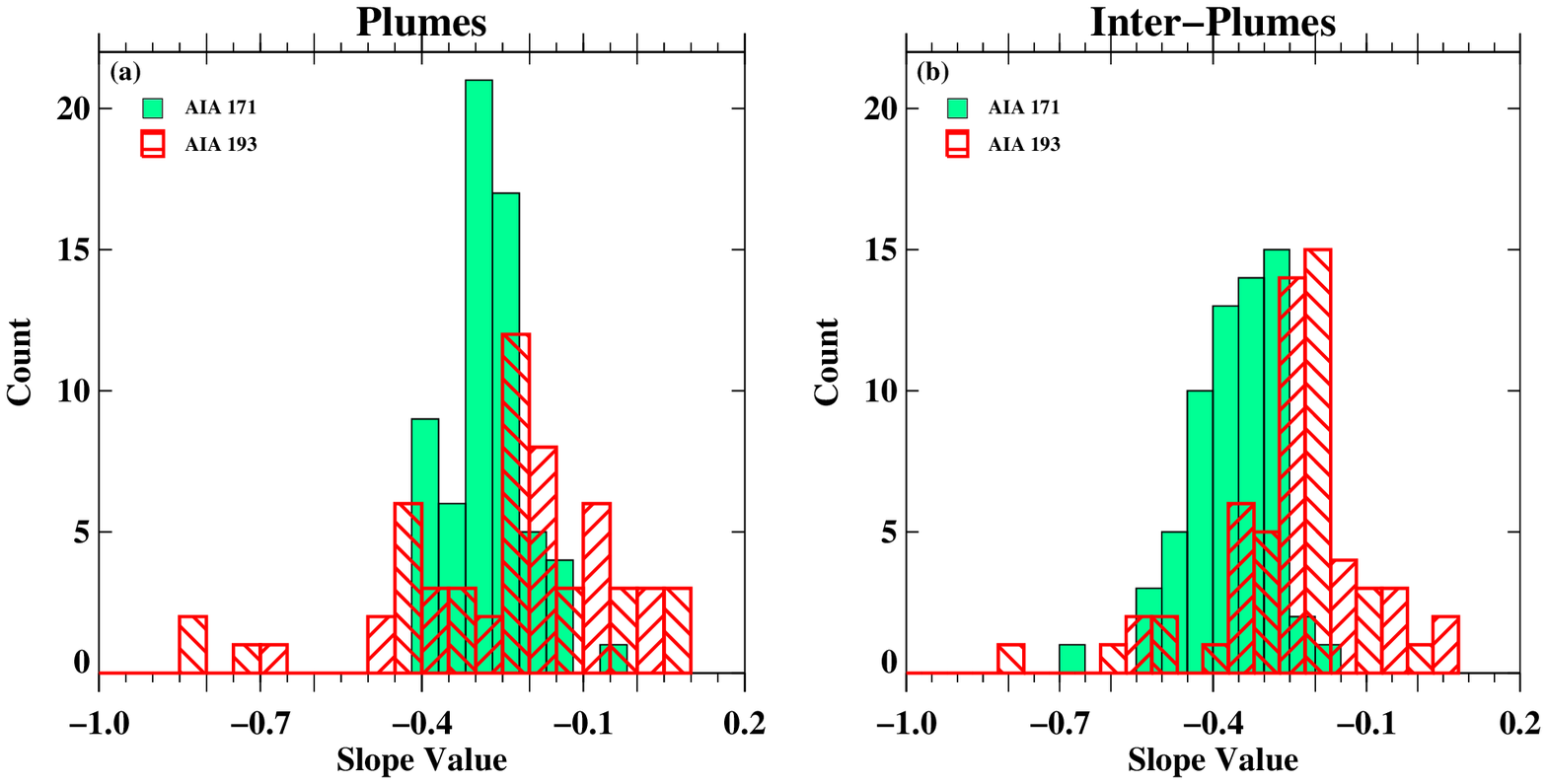}
\caption{Distributions of the slope values obtained from damping length-period relation similar to that derived in Figure~\ref{log_log}, but from individual datasets.} 
\label{indi_coeff}
\end{figure*}

At this point, it is also intriguing to find out the range of the slope values by examining individual datasets. We, therefore, regroup the periods and corresponding damping lengths obtained from individual datasets and fit them separately to obtain respective slope values. Figure~\ref{indi_coeff} displays the histograms of these values for plumes and interplumes. It appears the slope values in the 171~\r{A} passband are narrowly distributed around -0.3 for plumes and skewed towards more negative values in interplumes. The corresponding values in the 193~\r{A} passband show a relatively wider distribution even with a few cases of positive (but close to zero) values. A majority of the cases still show negative slopes but the peaks are shifted towards the less negative side compared to that in the 171~\r{A} passband. In order to check the similarities between the distributions shown in Figure~\ref{indi_coeff}, we perform two nonparametric tests, namely Kolmogorov-Smirnov (K-S) test and the Mood test. The K-S test quantifies the distribution equivalence whereas the Mood test checks for the equality of scale \citep{daniel1990applied}. All these tests have been performed using the statistical analysis software `R' \citep{feigelson2012modern}. Both the K-S test results (D$_{plume}$=0.4, D$_{interplume}$=0.6 with P$<$0.001) and the Mood test results (P$<$0.001) indicate that the two distributions (panel(a) and panel(b) of Figure~\ref{indi_coeff}) are statistically different.

Let us now focus on the physical interpretation of the observed frequency dependence. The shorter damping lengths at the longer periods require the damping mechanism to be more efficient at low frequencies. Although there are several physical processes such as thermal conduction, compressive viscosity, radiation loss, gravitational stratification, magnetic field divergence, phase mixing, mode coupling etc., that could contribute to the damping of slow waves \citep{2003A&A...408..755D,2004A&A...415..705D,2004A&A...425..741D}, most of them are independent on frequency while a few (e.g., thermal conduction, compressive viscosity) are rather more efficient at high frequencies according to one-dimensional linear MHD wave theory \citep[see Table 1 in][]{2014ApJ...789..118K}. One possible scenario is that, a linear theory may not be a good scheme to describe slow waves in polar regions. In fact, \citet{2000ApJ...533.1071O} have used 1-D and 2-D MHD codes to study the nonlinear effects associated with slow wave damping in case of polar plumes and interplumes. These authors found that because of lesser plasma density, a stronger dissipation, in the non-linear damping regime, is expected in the interplumes as compared to the plumes. However, from our observations (see Figure~\ref{log_log}) we do not find any significant difference in damping length values in these two regions. It is also interesting to note that the slope values too are not much different despite the plasma parameters like density, temperature (and their gradients) being quite different in plumes and in surrounding interplumes \citep{2006A&A...455..697W,2011A&ARv..19...35W}. Furthermore, the fundamental mechanism through which the nonlinearly steepened slow waves are expected to dissipate is compressive viscosity which is mainly effective at high frequencies. So it is not yet clear what mechanism causes this unusual frequency dependence in the damping of slow waves. Perhaps, we need a full three-dimensional description of slow waves to better understand their dissipation characteristics in polar regions.

\subsection{Gaussian decay of amplitude}
It has been shown that the amplitude of damping kink waves could be better described with a Gaussian profile rather than with the regularly used exponential profile \citep{2012A&A...539A..37P, 2016A&A...585L...6P}. The strong damping observed in these waves is believed to be due to mode coupling \citep{2012A&A...539A..37P, 2013A&A...551A..39H} which is frequency dependent \citep{2010A&A...524A..23T}. Although the equivalent theory and simulations are not available yet for slow waves, it is possible to check if the damping characteristics of slow waves show similar signatures. Indeed, \citet{2014ApJ...789..118K} have shown (using a single example from a coronal loop) that for slow waves with longer periods, the amplitude decay is better approximated to a Gaussian function as compared to an exponential one. Using the large statistics available in this study, we explore this behavior by refitting the amplitude profiles (see Figure~\ref{context_image}e) of all the detected periods with a Gaussian function (defined as $\mathrm{A(x)=A_0\exp^{-x^{2}/L_d^{2}}+C}$). We note the corresponding $\chi^2$ values as a measure of goodness of fit and compare them with analogous values obtained from the exponential function defined in section~\ref{method}. It turns out that in only about 4\% of the cases the Gaussian model fares better than the exponential model for the data from the 171~\r{A} passband. This fraction is even less ($<$3\%) for the data from the 193~\r{A} passband. However, interestingly, the periodicities in all those cases were $\ge$20 min. To check for the robustness of the above results, we also fit the damping curves using the `maximum likelihood estimation' (MLE) method and calculate the Akaike's `An Information Criterion' (AIC) to choose between the models (the smaller the AIC, the better the fit). We use the `mle'( includes `optim' also) procedure in $\mathrm{R}$ to compute the parameters. Similar results were obtained confirming our finding that the amplitude decay in a very few cases (with periods $\ge$ 20 min) appears to be a Gaussian-like rather than an exponential one. 

 In any case, at this point, we are inclined to believe that an exponential decay model is good enough to describe the damping of slow waves.

\begin{figure*}[!htb]
\centering
\includegraphics[width=0.98\textwidth]{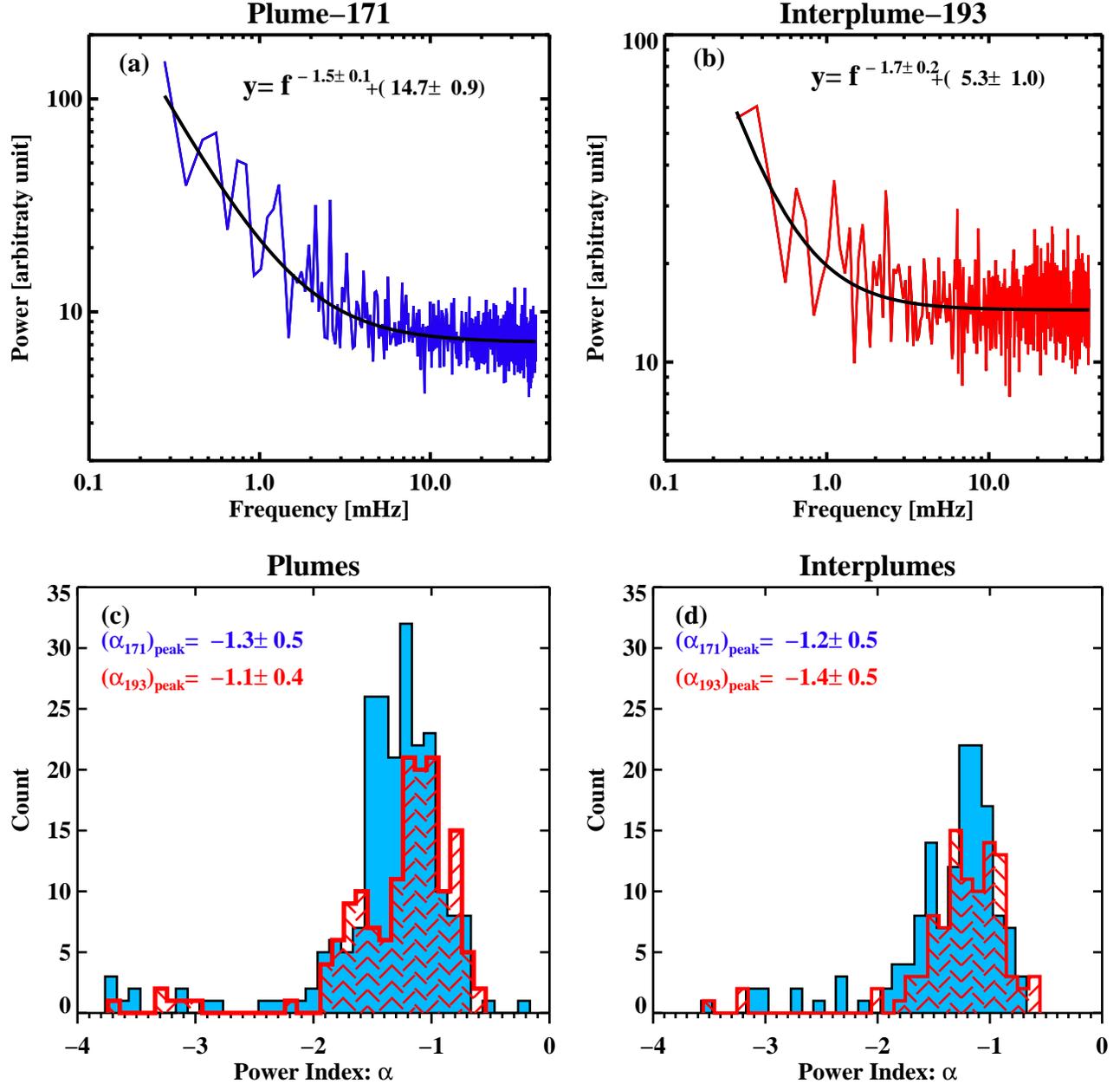}
\caption{Panels (a-b) show two representative Fourier power spectra derived from the two AIA channels. The solid black curves represent the power-law fits, the indices of which are listed in the plot. The distributions of the power-law index ($\alpha$), derived from all the datasets, are shown in panels (c-d) for different structures and passbands. Respective peak $\alpha$ values are printed on the panels.} 
\label{turbulence_plot}
\end{figure*}

\subsection{The power-law index}
In the previous sections we have discussed how the power at distinct peak periods decreases with distance as a consequence of the damping of slow waves and compared them to study the frequency-dependence. However, it is also important to compare the relative levels of power at the full spectrum of frequencies with respect to each other particularly in the context of understanding MHD wave turbulence \citep{2015RSPTA.37340148C}. One of the common ways to perform this is by fitting a power-law function ($\mathrm{f^{\alpha}}$) to the power spectra and noting the power-law index, $\alpha$ \citep{2017arXiv170702448B,2016A&A...592A.153K}. Such studies have been done in the past for different solar features: in chromospheric network and fibrils \citep{2008ApJ...683L.207R}; in active regions \citep{2014A&A...563A...8A}; in polar plumes \citep{2014A&A...568A..96G}; in sunspot, moss and quite Sun \citep{2015ApJ...798....1I}. In fact, \citet{2014A&A...568A..96G} studied the power-law behavior only at six locations in the polar region and found that it resembles a Kolmogorov-turbulence like behavior \citep{1941DoSSR..30..301K}. Utilizing the large database, in this study, we derive the power-law indices from intensities at several plume, interplume structures corresponding to the two AIA passbands.
 A pixel location at about 30$^{\prime\prime}$ away from the solar limb (guided by our previously selected artificial slits) is chosen for each plume, interplume structure.
In order to minimize the effect of noise, we perform an averaging, during the power spectra generation processed, on a  5$\times$5 pixel$^2$ area surrounding this location. This averaging, in this case, is done in Fourier domain i.e light curves from each of these pixels (within the mentioned 5$\times$5 area) are subjected to Fourier analysis to get the corresponding power spectrum and all these power spectra are then averaged to generate the final power spectra which is then fitted with a power-law function to get the power-law index.
 The locations closer to the limb are selected in order to get the power spectrum in full period range \citep{2014A&A...568A..96G}. While fitting the spectrum, we use statistical weights (i.e $1/\sqrt{\mathrm{y}}$) to minimize the effect of spurious power fluctuations around the higher frequencies. All the spectra have been fitted with a function $\mathrm{Y=f^{-\alpha}+C}$ where the constant term ($\mathrm{C}$) represents the frequency-independent 'white-noise' tail. Panels (a-b) of Figure~\ref{turbulence_plot} show two representative final Fourier power spectra (in log-log scale) obtained from a plume and an interplume structure. The overplotted black solid curves represent the fitted functions. The obtained $\mathrm{\alpha}$ values are also listed in the plot. We fit the spectra only upto 0.27$\mathrm{mHz}$ (60 min) keeping in mind that the duration, of every individual dataset, is 180 min.
 It may be noted that the $\alpha$ values derived in this way may have some influence from the chosen spatial binning of the data due to the autocorrelation and non-Gaussian noise.

We also constructed histograms of all the obtained $\alpha$ values to compare their distributions across different structures and passbands which are shown in panels (c-d) of Figure~\ref{turbulence_plot}.The peak values from the respective data are listed in the plot. An immediate observation reveals that all these distributions (in different structures and passbands) are fairly similar despite the different plasma parameters and magnetic field values in these regions. As we note, though our distributions have peaks around $\mathrm{\alpha}$$\approx$$-$1.2 but there are large number of occurrences close to $\mathrm{\alpha}$=$-$1.6. Such an $\mathrm{\alpha}$ value is probably a signature of the classical Kolmogorov like ($\mathrm{\alpha}$=$-$5/3) turbulence \citep{1941DoSSR..30..301K}. On the other hand, \citet{1674-4527-16-6-008} has shown that the peak at $\mathrm{\alpha}$$\approx$$-$1.2 can also be interpreted as a signature of the underlying generation mechanism which these authors attribute to the chromospheric spicules. The presence of both the scenario, as found in this study, demands an in-depth study of the power law index in these coronal structures (in fact, turbulence driven coronal heating models show that turbulence can play a major role in dissipating the wave energies into the surrounding medium \citep{2007ApJS..171..520C,2010ApJ...708L.116V}). Nonetheless, our study reveals a wider spectrum of $\mathrm{\alpha}$ values as opposed to the only `Kolmogorov-like turbulence' case as found by \citet{2014A&A...568A..96G}.


\section{Summary and Conclusions}
We have performed a statistical analysis of the frequency-dependent damping of slow waves using high resolution EUV imaging data. Previous case studies indicate anomalous behavior with longer wave periods damping faster than the shorter ones. In this study, we use a large sample of polar region data, to extract the exact frequency dependence, along with other useful properties. Below we summarize the main results obtained from this study:

$\bullet$ Intensity oscillations, in polar regions, are usually observed at longer periods. However, we find, the short-period (2--6 min) oscillations are at least as abundant as the long-period ($>$7 min) ones in both plume and interplume structures. The larger power available at the long-period oscillations makes them readily apparent suppressing the shorter periods.

$\bullet$ The damping length and the oscillation period are weakly related through an inverse power-law function as previously found. The slope, as obtained by fitting the logarithmic values of these quantities, is $-$0.33$\pm$0.05 for the plume and $-$0.38$\pm$0.05 for the interplume structures, in the 171~\r{A} passband. The corresponding values in the 193~\r{A} passband are $-$0.28$\pm$0.08 and $-$0.23$\pm$0.07 for the plume and the interplume structures, respectively. The negative slopes could not be explained within the one-dimensional linear MHD description of slow waves. Future theoretical and numerical studies with full three-dimensional MHD are required.

$\bullet$ A comparison of slopes obtained from individual datasets show that the values in the 193~\r{A} passband are mostly negative but lower indicating a shallow dependence. On the other hand, the data from interplumes in the 171~\r{A} passband show steeper dependence for a majority of the cases.

$\bullet$ The decay of amplitude could be well described with an exponential function. A Gaussian profile was found to serve better only in about 4\% of the cases, all of which correspond to longer periods ($>$20 min).

$\bullet$ Lastly, the Fourier power spectra near the limb show a power-law distribution with a peak close to $\alpha$=$-$1.2, suggesting a possible connection between the lower atmosphere transients with the generation of the observed coronal slow waves \citep{1674-4527-16-6-008}. However, significant number of cases around $\alpha$=$-$5/3 perhaps indicate the physical nature of the underlying turbulence mechanism.


We hope these results will be useful to put some constraints on the numerical models of slow wave propagation and dissipation. Also, further studies are required to explore more about the effects of turbulence on the damping of slow waves.

\section{Acknowledgments}
The authors would like to thank the scientific editor and the referee for their valuable suggestions which helped us to improve the quality of the paper. The AIA data used here is courtesy of the SDO (NASA) and AIA consortium. The authors also want to acknowledge the Joint Science Operations Center (JSOC) for providing the download links of the AIA data. S.M also acknowledges Ms. Priyanka Rani for her help in reducing the AIA data.



\end{document}